\documentclass[floatfix,prl,twocolumn,superscriptaddress]{revtex4-1}

\usepackage{graphicx}
\usepackage{amsmath}
\usepackage{amsthm,amsmath,amsfonts,dsfont}
\usepackage{color}

\newcommand{\comment}[1]{}

\begin{document}

\title{High-density quantum sensing with dissipative first order transitions}

\author{Meghana Raghunandan}
\affiliation{Institut f\"ur Theoretische Physik, Leibniz Universit\"at Hannover, Appelstra{\ss}e 2, 30167 Hannover, Germany}
\author{J\"org Wrachtrup}
\affiliation{3.~Physikalisches Institut, Universit\"at Stuttgart, Pfaffenwaldring 57, 70569 Stuttgart}
\author{Hendrik Weimer}
\email{hweimer@itp.uni-hannover.de}
\affiliation{Institut f\"ur Theoretische Physik, Leibniz Universit\"at Hannover, Appelstra{\ss}e 2, 30167 Hannover, Germany}

\begin{abstract}

  The sensing of external fields using quantum systems is a prime
  example of an emergent quantum technology. Generically, the
  sensitivity of a quantum sensor consisting of $N$ independent
  particles is proportional to $\sqrt{N}$. However, interactions
  invariably occuring at high densities lead to a breakdown of the
  assumption of independence between the particles, posing a severe
  challenge for quantum sensors operating at the nanoscale. Here, we
  show that interactions in quantum sensors can be transformed from a
  nuisance into an advantage when strong interactions trigger a
  dissipative phase transition in an open quantum system. We
  demonstrate this behavior by analyzing dissipative quantum sensors
  based upon nitrogen-vacancy defect centers in diamond. Using both a
  variational method and numerical simulation of the master equation
  describing the open quantum many-body system, we establish the
  existence of a dissipative first order transition that can be used
  for quantum sensing. We investigate the properties of this phase
  transition for two- and three-dimensional setups, demonstrating that
  the transition can be observed using current experimental
  technology. Finally, we show that quantum sensors based on
  dissipative phase transitions are particularly robust against
  imperfections such as disorder or decoherence, with the sensitivity
  of the sensor not being limited by the $T_2$ coherence time of the
  device. Our results can readily be applied to other applications in
  quantum sensing and quantum metrology where interactions are
  currently a limiting factor.

\end{abstract}


\maketitle


The challenges associated with quantum sensing within interacting
systems is particularly relevant for magnetic field sensing with
nitrogen-vacancy (NV) color centers in diamond, as strong magnetic
dipole interactions present a challenge to perform magnetometry at
high densities \cite{Rondin2014}. For NV centers, performing
magnetometry at high densities is particularly important, enabling to
study processes inside living cells \cite{Kuckso2013}. These challenges
imposed by interacting systems are not totally surprising, given that
the magnetic dipole moments of NV centers is what enables to measure
magnetic fields in the first place. Hence, the effect we are
addressing is quite generic and is also found in related applications;
for example, uncertainties caused by interactions are currently one of
the most important limiting factors for optical lattice clocks
\cite{Nicholson2015,Ushijima2015}.

\begin{figure*}
\begin{tabular}{p{0.4cm}p{8.5cm}p{0.4cm}p{8.25cm}}
\vspace{0pt}(a) & \vspace{0pt} \includegraphics[width=8.5cm]{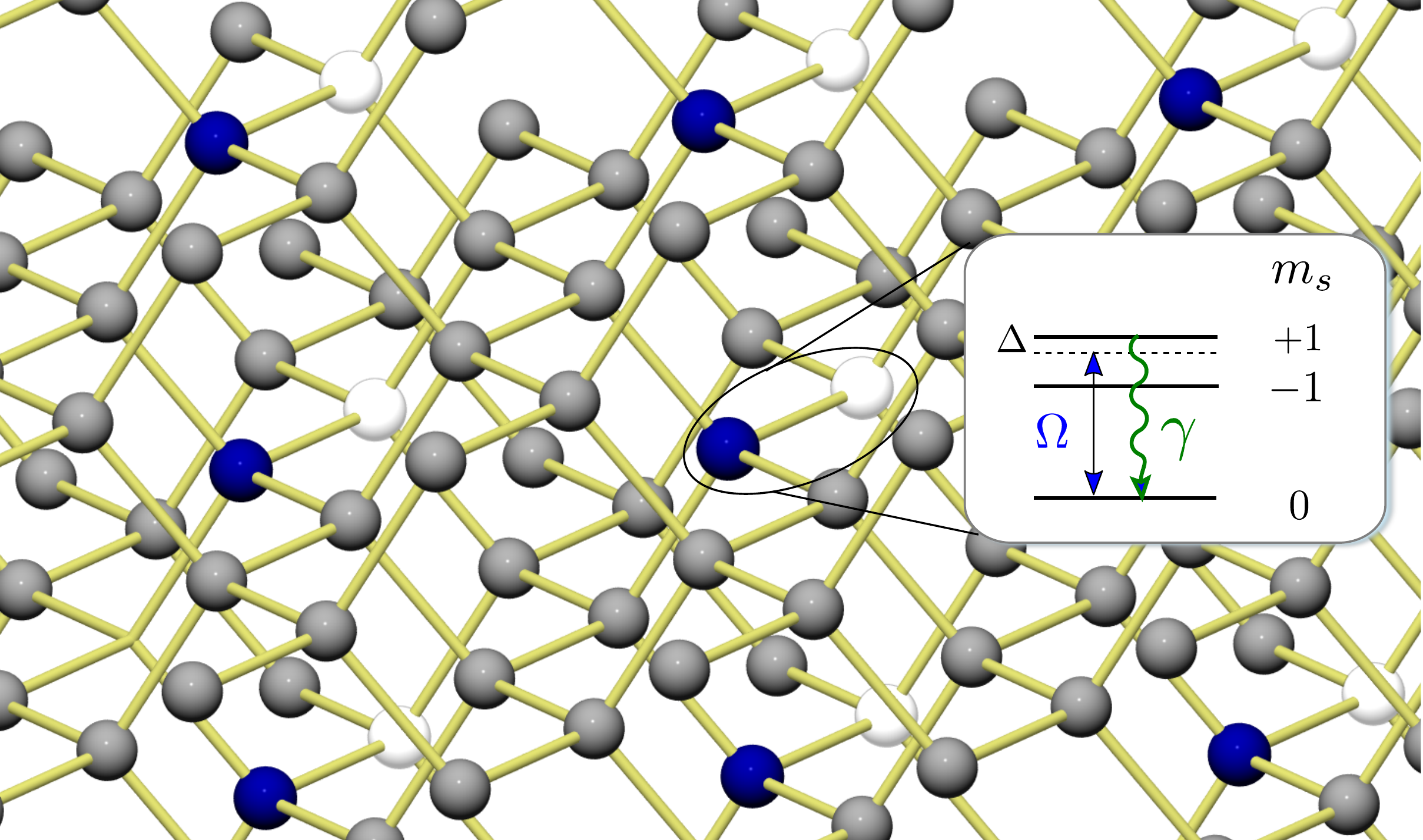} &\vspace{0pt} (b) &  \vspace{-0.25cm} \includegraphics[width=8.25cm]{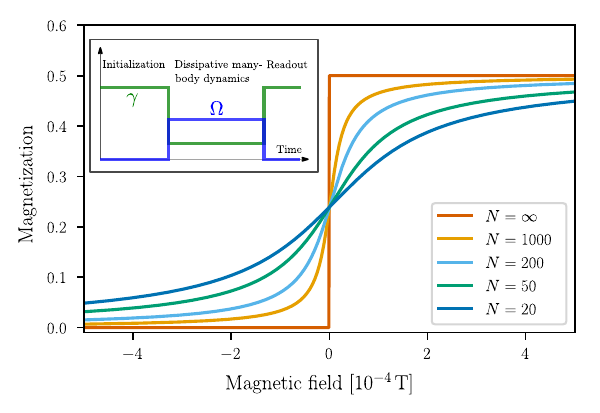}\\
\end{tabular}

\caption{Setup of the system for dissipative quantum sensing. (a)
  Many-body system of nitrogen-vacancy centers in diamond showing
  individual carbon atoms (gray) and nitrogen impurities (blue)
  accompanied by a vacancy site (white). The electronic ground state
  is a triplet state that is split by an external bias field. (b)
  Sketch of the dimensionless magnetization of the system across a
  first order phase transition. The response of the system strongly
  increases for larger system sizes. At the transition, the derivative
  of the magnetization is proportional to
  $\sqrt{N}$. \textcolor{black}{The inset shows the sensing protocol
    consisting of NV initialization, dissipative many-body dynamics,
    and readout of the NV spin state.}}

\label{fig:setup}

\end{figure*}

Building on the tremendous progress in controlling individual
\cite{Jelezko2004,Childress2006,
  Dutt2007,Taylor2008,Neumann2008,Maurer2012,Bar-Gill2013,Romach2015}
and interacting \cite{Zhu2011,Dolde2013,Choi2016} NV centers, combined
with the first studies investigating many-body effects
\cite{Yao2011,Weimer2012,Kessler2012,Weimer2013,Cai2013a,Albrecht2014,Choi2016a},
we consider large ensembles of microwave-driven NV centers interacting
via the magnetic dipole interaction. As an important ingredient, we
also incorporate optical pumping of the NV centers towards the $m_s
=0$ spin state, see Fig.~\ref{fig:setup}. Such driven-dissipative spin
systems are closely related to dissipative Ising models studied in
Rydberg gases \cite{Lee2011,Marcuzzi2014}, which exhibit a dissipative
first order liquid-gas transition at a critical strength of the
driving field \cite{Weimer2015,Weimer2015a,Kshetrimayum2016}, with the
first order transition line ending in a critical point belonging to
the Ising universality class \cite{Marcuzzi2014}. Crucially, the
susceptibility of the system diverges with the number of spins at the
transition point, showing a dramatic response of the system that can
be used for quantum sensing \cite{Gammelmark2011}. A key advantage of
turning to the steady state of a driven-dissipative system is that all
additional imperfections, such as disorder or decoherence, can be
integrated into the sensing process, meaning they only shift the
position of the transition without affecting its usefulness for
quantum sensing applications.

In this Letter, we demonstrate that the dissipative phase transition
is also present in the case of NV centers. Focusing first on the case
of two-dimensional arrays of NV centers, we perform a variational
analysis of the many-body system in thermodynamic limit. We compare
the variational results to wave-function Monte-Carlo simulations for
systems containing up to 20 spins, which to our knowledge, is the
largest number of spins treated so far in an open quantum many-body
systems while retaining the full Hilbert space.  We show that in
three-dimensional systems, the anisotropy of the dipole-dipole
interactions replaces the sharp phase transition by a smooth
crossover, however, the transition can easily be restored by applying
a magnetic field gradient of modest strength. Finally, we address the
role of additional imperfections and decoherence channels within the
setup and demonstrate that a finite $T_2$ coherence time does not
limit the sensitivity of the quantum sensor.

In our investigations, we consider a system of $N$ NV centers in a
lattice geometry. Such structures can be implemented using targeted
ion implantation at the nanometer scale
\cite{Scarabelli2016}. Furthermore, the NV centers can be
preferentially aligned along the axis of the external magnetic field
\cite{Karin2014}. We consider an effective two-level description of
the NV centers, where the $m_s=-1$ state is off-resonant with respect
to the microwave field, see Fig.~1, due to the external bias field
$B_0$. In the rotating frame of the driving, the Hamiltonian is of the
form
\begin{equation}
H = \frac{\hbar\Delta}{2}\sum\limits_i^N\sigma_z^{(i)} + \frac{\hbar\Omega}{2} \sum\limits_i^N\sigma_x^{(i)} + \sum\limits_{i<j}^NV_{ij},
\end{equation}
where $\Delta$ is the detuning from the electron spin resonance and
$\Omega$ is the Rabi frequency of the microwave driving. The
dipole-dipole interaction $V_{ij}$ is given by
\begin{eqnarray}
    V_{ij} &=& \left(1-3\cos^2\vartheta_{ij}\right)\frac{\mu^2}{|{\bf r}_i -{\bf r}_j|^3}\\
    &\times&
    \left\{\frac{1}{4}\left[1+\sigma_z^{(i)}\right]\left[1+\sigma_z^{(j)}\right] - \sigma_+^{(i)}\sigma_-^{(j)}-\sigma_-^{(i)}\sigma_+^{(j)}\right\},\nonumber
    \label{eq:V}
\end{eqnarray}
where ${\bf r}_i$ denotes the position of the NVs, $\mu$ indicates the magnetic
dipole moment, and $\vartheta_{ij}$ is the angle between the
NV axis and the vector connecting sites ${\bf r}_i$ and ${\bf r}_j$. We account for the optical pumping of the spins by considering a quantum master equation in Lindblad form,
\begin{equation}
  \frac{d}{dt}\rho = -\frac{i}{\hbar}[H,\rho] + \sum\limits_{i}^N \gamma \left(\sigma_-^{(i)}\rho \sigma_+^{(i)} - \frac{1}{2}\left\{\sigma_+^{(i)}\sigma_-^{(i)}, \rho\right\}\right),
\end{equation}
where $\gamma$ is the rate of the optical pumping process, which can
be controlled by the strength of the green pump laser. In all our
calculations, we assume the NV centers to be separated by $r =
5\,\mathrm{nm}$, and the optical pumping rate to be $\gamma =
1\,\mathrm{MHz}$. Unless we specifically investigate the response to
an additional magnetic field, we assume the driving to be on
resonance, i.e., $\Delta=0$.

As in the case of conventional NV sensors \cite{Rondin2014}, the
system is read out by the fluorescence signal from the NV centers in
the $m_s=1$ state. The only difference is that the dynamics of the
system does not follow a Ramsey sequence, but is governed by the
dissipative many-body dynamics of the quantum master equation.

\emph{Two-dimensional systems.---} We first turn to the analysis of
two-dimensional square lattices where the dipoles are oriented
perpendicular to the plane of the system. 
We also simplify the analysis by
considering only interactions between adjacent lattice sites; taking
the full long-range tail into account only slightly
modifies our results on a quantitative level, but the qualitative
findings will remain unchanged \footnote{See the Supplemental Material for a numerical investigation of the effects of disorder and the full dipolar interaction, as well as a detailed discussion of the sensitivity of the proposed quantum sensor.}
\begin{figure}[b!]

  \includegraphics{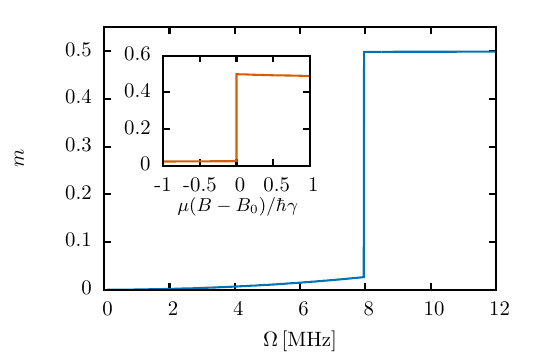}

  \caption{Variational solution for the steady state magnetization
    showing a first order phase transition. The phase transition can
    be triggered by varying either the Rabi frequency $\Omega$ or the
    external magnetic field (inset)($V = 2\pi\times
      400\,\mathrm{kHz}$, $\gamma = 1\,\mathrm{MHz}$).}

\label{fig:var2d}
\end{figure}
As a first step, we investigate the steady state of the quantum master
equation based on the variational principle for open quantum systems
\cite{Weimer2015}. Here, we use product states of the form $\rho =
\prod_i \rho_i$ as our variational basis, with $\rho_i$ being the
reduced density matrix at site $i$. Then, we find a first order
transition of the NV magnetization $m = \sum_i
(1+\langle\sigma_z^{(i)}\rangle)/(2N)$ in the driving strength
$\Omega$, see Fig.~\ref{fig:var2d}. This transition appears to be
closely related to what has been predicted for dissipative Ising
models discussed in the context of Rydberg gases, where the flip-flop
term of Eq.~(\ref{eq:V}) is absent \cite{Bernien2017}. Crucially, the
first order transition can also be triggered by modifying the external
magnetic field, allowing to use this transition for the sensing of
static fields.

\begin{figure}[t]
\includegraphics{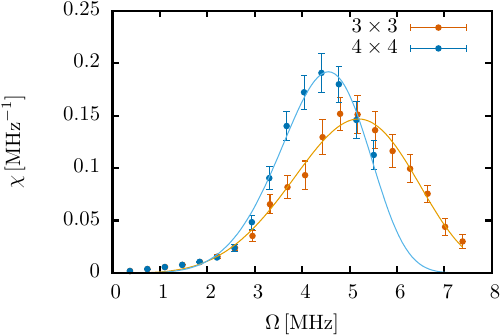}

\caption{Averaged results from 1000 wave-function Monte-Carlo
  trajectories showing the steady state susceptibility $\chi$ for
  $3\times 3$ and $4 \times 4$ geometries. The solid lines are fits to
  a Weibull distribution.}

\label{fig:mcsolve}
\end{figure}
\begin{figure}[t]
\includegraphics{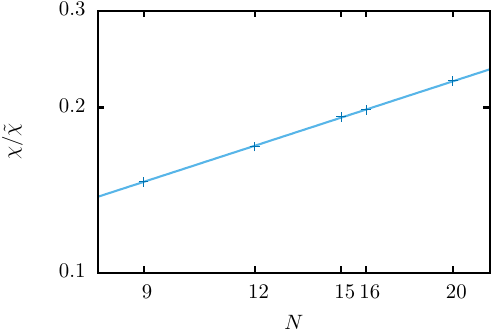}
\caption{Finite size scaling of the peak of the reduced susceptibility
  $\chi/\tilde{\chi}$. The collapse of the data onto a single line in
  the logarithmic plot shows that the susceptibility diverges with the
  system size $N$.  The solid line is an algebraic fit to the data.}

\label{fig:fss}

\end{figure}

\emph{Wave-function Monte-Carlo simulations.---} We perform numerical
simulations of the full quantum master equation for systems up to 20 spins. We
use the results from the simulations based on a wave-function
Monte-Carlo approach \cite{Johansson2013}, which we extended to a
massively parallelized version, to serve as a benchmark for our
variational analysis. In particular, we are interested in the
existence of the first order transition predicted by the variational
approach. For this, we investigate the magnetic susceptibility $\chi =
\partial m/\partial \Omega$, which diverges at a first order
transition. Figure~\ref{fig:mcsolve} shows the numerically obtained
susceptibility for different system sizes. Interestingly, we find that
the susceptibility data closely follows a Weibull distribution
$\chi(\Omega) \sim \Omega^{k-1}\exp[-(\Omega/\lambda)^k]$. We note
that the Weibull distribution has been discussed in the context of the
relaxation from metastable states, \textcolor{black}{with the parameter
  $k$ controlling their relative decay rates
  \cite{Maier1997,Adams2010}}. \textcolor{black}{Such metastable states}
also play an important role in dissipative Ising models
\cite{Rose2016,Letscher2016}. To investigate the scaling with the
number of spins $N$ in detail, we turn to a finite size scaling
analysis. \textcolor{black}{For this, we aim to describe the simulation
  results for the susceptibility peak in terms of a scaling function,
  from which we can extract how the susceptiblity peak changes with
  $N$. Here, we also include anisotropic geometries to be able to
  treat larger system sizes up to 20 spins. Our ansatz for the scaling
  function is given by}
\begin{equation}
  \chi = c N^\alpha \tilde{\chi}(\lambda),
  \label{eq:scaling}
\end{equation}
where $\lambda = N_x/N_y$ is the anisotropy given in terms of the
number of spins in the $x$ and $y$ direction, respectively, \textcolor{black}{while $c$
and $\alpha$ are numerical constants \cite{Binder1989}. Crucially,
when the exponent $\alpha$ is found to be positive, the susceptibility
diverges with $N$, signalling the presence of a first order
transition. The reduced scaling
function $\tilde{\chi}$ captures the effects of anisotropic system sizes and}
must satisfy the conditions
$\tilde{\chi}(\lambda) = \tilde{\chi}(1/\lambda)$ and $\tilde{\chi}(1)
= 1$. Consequently, we can perform a series expansion according to
$\tilde{\chi}(\lambda) = 1 + d\left[\log \lambda\right]^2 + O(\log [\lambda]^4)$,
which we can truncate for not too large anisotropies. $d$ is another
numerical constant that can be determined from fitting to the
simulation data. \textcolor{black}{Then, the reduced susceptibility
  $\chi/\tilde{\chi}=cN^\alpha$ should be given by a simple algebraic
  function. Fig.~\ref{fig:fss} demonstrates that this is indeed the
  case, showing that the ansatz of Eq.~(\ref{eq:scaling}) is correct,}
confirming the existence of the first order transition.
The observed exponent $\alpha = 0.527\pm 0.006$ shows that the system
exhibits basically the same scaling of the sensitivity for quantum
sensing as a noninteracting ensemble ($\alpha = 1/2$).

\emph{Three-dimensional systems.---} As the next step, we will study
the properties of the system in three spatial dimensions. This will be
especially important as controlling the implantation depth of the NV
centers will be particularly challenging, making it natural to focus
on effectively three-dimensional setups. Here, we turn to a
three-dimensional cubic lattice to investigate the consequences. In
particular, the anisotropy of the dipole-dipole interaction will now
play an important role. Crucially, the dipole-dipole interaction
vanishes when integrated over the full solid angle, as the
ferromagnetic and antiferromagnetic contributions exactly cancel each
other. To capture this property in our nearest-neighbor model, we set
the interaction energy within the plane of the dipoles to
$V=\mu^2/r^3$ and to $-2V$ in the third direction.

\begin{figure}[t]
\includegraphics{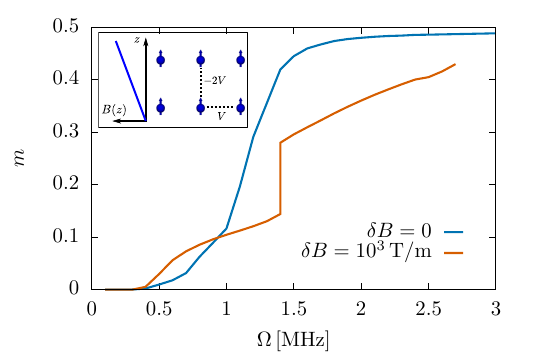}

\caption{Variational solution for the magnetization of the central
  layer in a three-dimensional system. The first order transition is
  replaced by a smooth crossover. Using a magnetic field gradient
  \textcolor{black}{(inset)}, the phase transition can be restored.}

\label{fig:var3d}
\end{figure}

In three dimensions, the system sizes are prohibitively large for
wave-function Monte-Carlo simulations. Therefore, we restrict our
analysis to the variational approach, noting that in larger
dimensions, the variational solution is even closer to the exact
steady state \cite{Weimer2017}. Here, we consider a system consisting
of three two-dimensional layers, with the central layer being at the
zero point of the magnetic field gradient. Within the variational
analysis, we find that the other two layers are almost completely
polarized in the $m_s=0$ state, i.e., adding additional layers will
not change the results. Additionally, we find that the anisotropy of
the dipole-dipole interaction replaces the first order transition by a
smooth crossover, see Fig.~\ref{fig:var3d}. Nevertheless, it is
possible to recover the transition by applying a magnetic field
gradient along the NV axis, effectively breaking the symmetry of the
dipolar interaction. The first order transition appears already for
quite modest field gradients on the order of $\delta B = 10^3
\,\mathrm{T/m}$, which are readily achievable in experiments. For
larger values of the gradient, the first order jump in the
magnetization will be even more pronounced, \textcolor{black}{eventually
  recovering the 2D results for very strong gradients.} This
underlines the usefulness of dissipative quantum sensing even for
three-dimensional systems.

\emph{Decoherence and other imperfections.---} So far, our analysis
has been based on a rather idealized setup. In any real diamond
sample, there will be several sources of imperfections related to
decoherence or to disorder from imperfect positioning of the NV
centers. First, we want to point that disorder in the NV interaction
energies or missing sites in the lattice due to off-axis NV centers
are not going to play an important role. Crucially, these
imperfections only affect the strengths of the coupling constants, but
cannot reverse their signs. From the analysis of random-bond Ising
models \cite{Binder1986}, it is known that the underlying phase
transition is robust against such a type of disorder, which is consistent with our numerical simulations for disorder in the system \cite{Note1}. This leaves
decoherence processes caused by residual nitrogen impurities and
$^{13}$C nuclear spins as the dominant challenge. Hence, we
investigate in detail how a limited $T_2$ time caused by these
decoherence processes will affect the performance of the dissipative
quantum sensor.

\begin{figure}[t]
\includegraphics{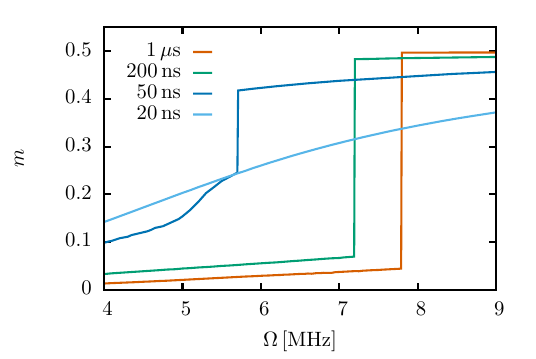}

\caption{Consequences of additional decoherence channels for different $T_2$ times. The first order transition is particularly robust to these additional decoherence processes, even when their associated rates become larger than the strength of the dipole-dipole interaction between the NV centers. Only for very short $T_2$ times, the phase transition is replaced by a smooth crossover.
 }

\label{fig:deco}

\end{figure}

Within the variational analysis, we add additional jump operators $c_i
= \sqrt{1/T_2}\,\sigma_z^{(i)}$ to the quantum master
equation. Importantly, we find that the existence of the first order
transition is robust against quite strong decoherence rates, see
Fig.~\ref{fig:deco}. Crucially, the phase transition does not merely
survive in a regime where the decoherence is perturbatively small
compared to the dipole-dipole interaction, but even in a regime where
the decoherence rates are several times larger than the coherent
interaction strength, which amounts to $V=2\pi \times
400\,\mathrm{kHz}$ at a NV distance of $r=5\,\mathrm{nm}$ in our
case. We attribute this strong robustness against decoherence to the
steady state being an effective thermal \textcolor{black}{(but
  non-classical)} state \cite{Maghrebi2016}. Such a state is diagonal
in an appropriate energy eigenbasis, making it less susceptible to
decoherence processes. The additional decoherence also leads to a
shift in the transition point, requiring to characterize the coherence
properties of a device before employing it as a quantum sensor. For
more dilute NV samples, the global timescale of the system gets
reduced, leading to a stronger susceptibility to decoherence. E.g.,
for a NV distance of $r=11\,\mathrm{nm}$, the phase transition will be
replaced by a smooth crossover for $T_2$ at about $500\, \mathrm{ns}$
instead of $50\, \mathrm{ns}$. We would also like to point out that
both the first order transition and the robustness to decoherence
remain present without an external bias field $B_0$ \cite{Note1}.

Finally, we estimate the sensitivity of the dissipative quantum
sensor, which we can extract from the finite size scaling behavior of
the susceptibility, as the change in fluorescense from the NV centers
is proportional to the magnetic susceptibility
\cite{Rondin2014}. Within our wave-function Monte-Carlo simulations,
we find that the susceptibility for DC fields $\chi_{DC} =\partial
m/\partial B$ shows very similar behavior as the AC susceptibility
\cite{Note1}, so the DC and AC sensitivity of the sensor are
essentially the same. This is very different in NV magnetometry using
noninteracting ensembles, as there the $T_2^*$ limited DC sensitivity
is generally worse than the $T_2$ limited AC sensitivity since DC
sensing does not allow for dynamical decoupling techniques
\cite{Rondin2014}. As our proposed sensor is not limited by $T_2^*$,
we expect our approach to be particularly useful for sensing DC
fields.  We can infer the sensitivity of the dissipative sensor to be
$\eta \approx 3\,\mathrm{nT}\,\mathrm{Hz}^{-1/2}$ for $N=10^3$ spins
and $\eta \approx 300\,\mathrm{fT}\,\mathrm{Hz}^{-1/2}$ for
$N=10^{11}$ \cite{Note1}. This sensitivity is approximately a factor
of three improvement over what has been recently demonstrated using
large ensembles of noninteracting NV centers \cite{Wolf2015}, while at
the same time offering a much smaller sensor size. Additionally, we
would like to stress that our dissipative quantum sensor can tolerate
large decoherence rates and operate at very small sensor sizes. These
unique features makes it extremely promising to use dissipative NV
sensors in \textcolor{black}{NV-rich na\-no\-di\-a\-monds
  \cite{Su2013}}, e.g., for the investigation of biological processes
inside living cells.

In summary, we have established a quantum sensing protocol based on
dissipative phase transitions. We have demonstrated the usefulness of
our approach for quantum sensing with nitrogen-vacancy defect centers
in diamond, finding a strong resilience of our protocol against
decoherence processes. Finally, our protocol does not depend on the
microscopic details of the sensing process, allowing for an immediate
transfer to other applications in quantum sensing (see \cite{Wade2017}
for a concrete example) and quantum metrology.

\begin{acknowledgments}
We thank D.~D.~B.~Rao for fruitful discussions. This work was funded by the Volkswagen Foundation and the DFG within SFB 1227 (DQ-mat).
\end{acknowledgments}

\bibliographystyle{aip}
\bibliography{bib}


\end{document}




%
%

\title{Supplemental Material for ``High-density quantum sensing with dissipative first order transitions''}

\author{Meghana Raghunandan}
\affiliation{Institut f\"ur Theoretische Physik, Leibniz Universit\"at Hannover, Appelstra{\ss}e 2, 30167 Hannover, Germany}
\author{J\"org Wrachtrup}
\affiliation{3.~Physikalisches Institut, Universit\"at Stuttgart, Pfaffenwaldring 57, 70569 Stuttgart}
\author{Hendrik Weimer}
\affiliation{Institut f\"ur Theoretische Physik, Leibniz Universit\"at Hannover, Appelstra{\ss}e 2, 30167 Hannover, Germany}



\maketitle

\section{Effect of long-range interactions and disorder}
We analyze the effects of long-range dipolar interactions and disorder caused by vacancies in our system. We compare the following three cases to the simplified case of nearest neighbour interaction for a $3 \times 3$ geometry: i. full long-range dipolar interactions, ii. disorder caused by a vacany in the lattice with only nearest neighbour interactions, and, iii. disorder caused by a vacany in lattice with full long-range dipole interactions. 

\begin{figure}[h]
\includegraphics{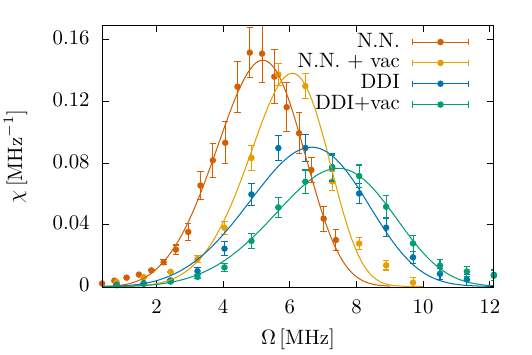}

\caption{Averaged results from 1000 wave-function Monte-Carlo trajectories showing the steady state susceptibility $\chi$ for different cases of $3\times 3$ geometry. The orange and yellow lines are for the cases of nearest neighbour interaction (N.N.) and the blue and green lines are for the cases of long-range dipole intearctions (DDI). The yellow and green lines include a vacancy (vac) in the lattice.}

\label{fig:chi2}

\end{figure}

In order for the geometries to have comparable interaction energies, we normalize the interaction strength $V_{ij}$ by calculating the total strength for each of the geometries and diving them with that of the nearest-neighbour case. We then run the wave-function Monte-Carlo simulations as discussed earlier and the results are shown in Fig.~\ref{fig:chi2}. We see that including the long-range interaction and vacancy disorders modify our results only quantitaively and the qualitative findings remain the same.

\section{Sensing in the absence of an external bias field}

\begin{figure}[h!t]

  \includegraphics{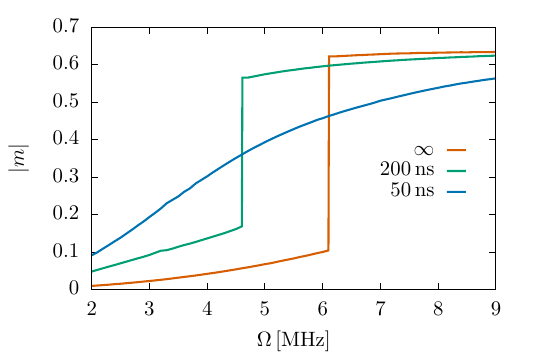}

  \caption{Absolute value of the magnetization of the NV centers in
    the absence of an external magnetic field. Without additional
    decohrence processes ($\infty$) or associated $T_2$ times on the
    order of several hundred nanoseconds, the first order transition
    required for sensing remains present.}

    \label{fig:spin1}

\end{figure}

In many applications, it is desirable to perform the sensing without
the presence of an external magnetic field $B_0$. In this case, the
$m_s=1$ and $m_s=-1$ states are degenerate, and it becomes necessary
to include the latter in the analysis. Using the variational principle
for open quantum systems \cite{Weimer2015}, we find a very similar
behavior of the absolute value of the magnetization of the system, see
Fig.~\ref{fig:spin1}, including again a remarkable robustness to
additional $T_2$ decoherence processes. Consequently, the dissipative
quantum sensor can also be operated with an external bias field.

\clearpage

\section{Magnetic susceptibility for DC fields}

\begin{figure}[ht]

  \includegraphics{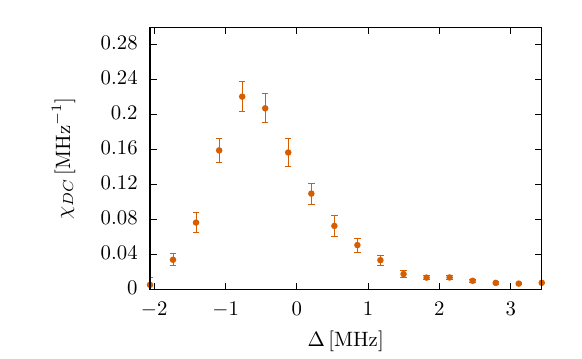}

  \caption{Averaged results from 1000 wave-function Monte-Carlo
  trajectories showing the steady state susceptibility $\chi_{DC}$ for a
  $3\times 3$ geometry.}

  \label{fig:dc}

\end{figure}

We calculate the magnetic susceptibility with respect to changes in
the magnetic field. As a variation of the magnetic field will shift
the resonance condition for the microwave driving of the NV centers,
we can capture the effect of a change in magnetic field with respect
to the external bias field $B_0$ in terms of a finite detuning
$\Delta$ in the Hamiltonian (1). Fig.~\ref{fig:dc} shows the magnetic
susceptibility $\chi_{DC} = \partial m/\partial \Delta$ as a function
of the detuning $\Delta$. As the peak of the susceptibility is only
slightly larger than the one for the AC susceptibility, the
sensitivities for AC and DC will be comparable.

\section{Sensitivity of the dissipative quantum sensor}

In contrast to non-interacting NV sensors, the measurement is not
limited by the photon shot noise of the individual NV centers, but
rather by the bimodal distribution of the system close to the first
order transition. In particular, the signal for the total
magnetization of the ensemble behaves as $N(m_0 + \chi_{DC} B)$, where
$B$ is the external field to be measured. The noise from the bimodal
statistics is given in terms of its standard deviation by $N/2\delta
m$, where $\delta m$ is the jump in magnetization at the transition in
the thermodynamic limit. From the numerical simulations, we see that
the susceptibility behaves essentially like $\sqrt{N}$ close to the
transition, leading to an overall $N^{-1/2}$ scaling of the
sensitivity. The sensitivity is then given by
\begin{equation}
\eta_{DC} = \frac{\delta m}{2 \chi_{DC} \sqrt{\nu}},
\end{equation}
where $\nu$ is the rate of the measurement. For $\delta m = 0.3$, $\nu
= 1\,\mathrm{MHz}$, we obtain $\eta \approx
3\,\mathrm{nT}\,\mathrm{Hz}^{-1/2}$ for $N=10^3$ spins and $\eta
\approx 300\,\mathrm{fT}\,\mathrm{Hz}^{-1/2}$ for $N=10^{11}$.


\newsavebox\myemptybib

\savebox\myemptybib{\parbox{\textwidth}{}}